\documentclass[aps,prl,twocolumn,superscriptaddress]{revtex4-1}

\usepackage{amsfonts}
\usepackage{amssymb}
\usepackage{amsmath}
\usepackage{graphicx}
\usepackage{natbib}
\setcitestyle{numbers,square}

\usepackage[usenames,dvipsnames]{color}
\usepackage{soul}

\usepackage{hyperref}

\renewcommand{\vec}[1]{\boldsymbol{#1}}

\newcommand{\ie}{{\it i.e.}}

\newcommand{\etal}{{\it et al.}}

\usepackage{multirow}

\begin{document}


\title{Field-angle dependence of sound velocity in the Weyl semimetal TaAs}


\author{F.~Lalibert\'e}
\author{F.~B\'elanger}
\affiliation{Institut Quantique and D\'epartement de physique, Universit\'e de Sherbrooke, Sherbrooke, Qu\'ebec J1K 2R1, Canada}

\author{N.~L.~Nair}
\author{J.~G.~Analytis}
\affiliation{Department of Physics, University of California Berkeley, Berkeley, California 94720, USA}

\author{M.-E.~Boulanger}
\author{M.~Dion}
\author{L.~Taillefer}
\affiliation{Institut Quantique and D\'epartement de physique, Universit\'e de Sherbrooke, Sherbrooke, Qu\'ebec J1K 2R1, Canada}

\author{J.~A.~Quilliam}
\email[]{jeffrey.quilliam@usherbrooke.ca}
\affiliation{Institut Quantique and D\'epartement de physique, Universit\'e de Sherbrooke, Sherbrooke, Qu\'ebec J1K 2R1, Canada}

\date{\today}

\begin{abstract}

The elastic modulus $c_{44}$ of a single crystal of the Weyl semimetal TaAs was investigated by measuring relative changes in the sound velocity under application of a magnetic field up to 10~T. Using an ultrasonic pulsed-echo technique, we studied the shear response of the crystal when the angle between the sound wave propagation and the magnetic field is changed. We observe a broken tetragonal symmetry at fields above 6 T, an anisotropy that is likely related to a longitudinal negative magnetoresistance and therefore might provide evidence of the chiral anomaly, one of the main topological signatures of this class of materials. We also observe quantum oscillations in the sound velocity whose frequencies vary with magnetic field orientation.  A fan diagram of Landau level indices reveals topological and trivial Berry phases, depending on the field orientation, indicating a sensitivity to different Fermi surface pockets that do or do not enclose Weyl nodes respectively. 

\end{abstract}


\maketitle


\section{Introduction}

In recent years, numerous experiments have attempted to show the existence of the
chiral anomaly in Weyl semimetals. One of the most cited pieces of evidence of the anomalous Adler-Bell-Jackiw contribution is undoubtedly the longitudinal negative magnetoresistance (LNMR)~\cite{arnold2016b,hirschberger2016,huang2015,zhang2016,niemann2017,li2015}
predicted to occur when  electric and magnetic fields are parallel, creating a charge transfer between two Weyl nodes of opposite chirality~\cite{nielsen1983,hosur2013}. While observations of LNMR could indeed be signatures of the chiral anomaly, some doubts have been raised with the suggestion that they could also be the effect of a non-uniform current distribution~\cite{dosReis2016a,ramshaw2017}. Other claims of non-trivial topological effects in Weyl semimetals include high magnetic field studies in the quantum limit of transport and thermodynamic properties~\cite{moll2016b,ramshaw2017}, optical conductivity measurements~\cite{noh2017}, Fermi surface topology studies via quantum oscillations~\cite{moll2016a,dosReis2016b,klotz2016} and Angle-resolved photoemission spectroscopy (ARPES)~\cite{lv2015,xu2015,yang2015}. In many cases, the electronic properties associated with the chiral anomaly are diluted with those of the topologically trivial quasiparticles, making the analysis more complex.

Here, we report a new approach to revealing the chiral anomaly in the canonic Weyl semimetal TaAs. Given the coupling between the lattice and the conduction electrons, we have measured the sound velocity in the presence of a magnetic field for a transverse acoustic mode. Considering that phonons are generally insensitive to magnetic field and in the absence of any field-induced phase transitions, the observation of a dependence of sound velocity or attenuation on field strength or orientation may be attributed to a coupling between acoustic phonons and conduction electrons. More precisely, the shear strain waves propagating in this material give rise to an oscillating electric field oriented along the direction of propagation, $\vec{q}$, via piezoelectricity. The discovery of a field-induced anisotropy, for example a change in sound velocity related to $\vec{q}\cdot\vec{B} \propto \vec{E}\cdot\vec{B}$, may provide a promising measure of the chiral anomaly. Indeed an increase in sound \emph{attenuation} $\Delta \Gamma \propto |\vec{B}|\cos^2\varphi$, where $\varphi$ is the angle between sound wave propagation $\vec{q}$ and $\vec{B}$, was predicted independently in two theoretical papers~\cite{spivak2016,pikulin2016}. More recently, Rinkel \emph{et al.} also predicted a decrease in sound velocity as the magnetic field is increased in the quantum limit where only the chiral Landau level (LL) remains active~\cite{rinkel2019}. However, the predicted angular dependence is rather non-trivial, with a constant decrease in velocity for all angles except a narrow window around $\varphi = \pi/2$. While this effect is directly related to the LNMR that is expected in transport measurements, here it is the dynamical conductivity at the sound-wave frequency that is relevant. Effects of the chiral anomaly on optical phonons have also been considered theoretically~\cite{song2016,rinkel2017} and while one might expect stronger effets where optical phonon and plasmon frequencies are matched, sound velocity measurements can be performed to extremely high resolution ($< 1$ ppm) permitting the detection of rather subtle effects on acoustic phonons. 
 

 \begin{figure*}
\begin{center}
 \includegraphics[width=6.5in]{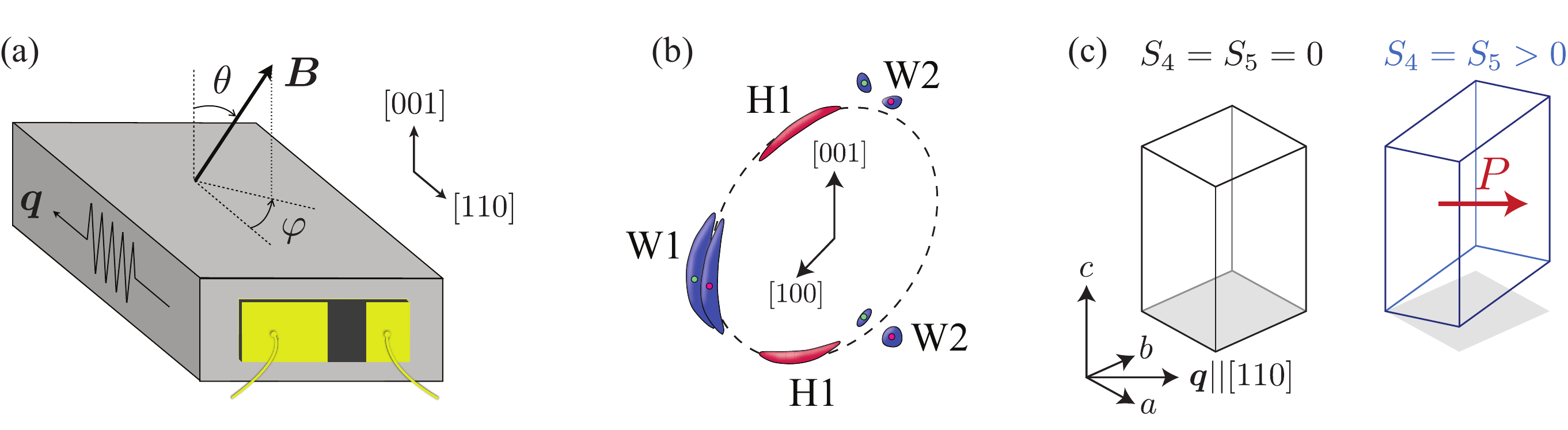}
 \end{center}
 \caption{(a) Schematic of the sample showing the definition of the azimuthal angle $\varphi$
 and polar angle $\theta$ of the magnetic field $\vec{B}$ with respect to the sound propagation direction $\vec{q}$.  (b) A sketch of a part of the Fermi surface reported in Ref.~\cite{arnold2016a} showing two types of Weyl pockets (W1 and W2) and trivial hole pockets (H1). Green and red dots indicate opposite chiralities of the Weyl nodes. Here we show just the part of the Fermi surface oriented along $[100]$. There are also equivalent sections of the Fermi surface oriented along $[010]$, $[\bar{1}00]$ and $[0\bar{1}0]$, thereby respecting the tetragonal symmetry of the lattice. (c) The unstrained (left) and strained (right) unit cell. With propagation vector $\vec{q} || [110]$ and displacement $\vec{u}|| [001]$, equal $S_4$ and $S_5$ components of the strain tensor are generated. The $d_{15} = d_{24}$ components of the piezoelectric tensor then result in a polarization $\vec{P} || \vec{q}$.
 \label{sample}}
 \end{figure*}


Our measurements of the $c_{44}$ elastic constant reveal rich behavior as a function of magnetic field amplitude and orientation. Two important results can be extracted from our measurements. First, we discover quantum oscillations (QOs) originating from small Fermi surfaces, in agreement with previous Shubnikov-de Haas (SdH) and de Haas-van Alphen (dHvA) measurements~\cite{arnold2016a}. We will show that these oscillations can be used to determine the Berry phase of the various Fermi pockets of TaAs and thereby identify those of a topological nature (containing a Weyl node) and those that are topologically trivial. Second, we show a breaking of tetragonal symmetry for magnetic fields above $\sim 5$ T as we tune the angle between sound-wave propagation and the magnetic field within the $ab$-plane,
an effect that we suggest, given theoretical predictions~\cite{rinkel2019}, may be a consequence of the chiral anomaly.



\section{Experimental method}

TaAs single crystals were grown by chemical vapor transport. Polycrystalline precursors were first synthesized using high purity Ta and As, ground and mixed with a 1:1 ratio. This material has a tetragonal structure with lattice parameters $a=b=3.44$~\AA~and $c=11.64$~\AA. The crystal was polished in order to obtain two opposite and parallel faces with mirror-like aspect, separated by $L=1.4$~mm in the [110] direction. Directions and angles are defined on the schematic drawing of the sample in Fig.~\ref{sample}(a). 

A pulsed-echo ultrasonic interferometer was used to measure the velocity and amplitude of transverse acoustic waves propagating along the $\vec{q} \parallel [110]$ direction, with polarization along $\vec{u}\parallel $[001]. In Voigt notation, the measured velocity is related to the $c_{44}$ element of the stiffness tensor via $v_{44} = \sqrt{c_{44}/\rho}$, where $\rho$ is the density of the material. Acoustic waves were generated with a LiNbO$_3$ piezoelectric transducer with fundamental frequency of $\sim 30$~MHz. The measurement technique consists of adjusting the frequency in order to maintain a constant phase of a given echo. Relative variations of velocity $\Delta v_{44}/v_{44}$ are then equal to relative changes in frequency, $\Delta f/f$. A rough absolute value of the velocity $v_{44} \simeq 2.8$~km/s, was obtained from the transit time between reflected echoes, in good agreement with calculations in Ref.~\cite{buckeridge2016}. 

This mode was chosen for several reasons. First, we have chosen the direction of sound wave propagation along $[110]$ to minimize trivial sources of anisotropy, as will be further discussed below. Second we have selected a mode that is piezoelectrically active and induces a dielectric polarization along the direction of sound wave propagation, allowing us to probe transport in the $ab$-plane. The longitudinal mode with $\vec{q}\parallel [110]$ would induce a dielectric polarization along $[001]$. Finally, we have selected a mode that probes a single element of the stiffness tensor, that is $C_{44}$~\cite{Note1}.

Here we assume that the coupling between this acoustic mode and conduction electrons is dominated by piezoelectricity, although coupling via a deformation potential could also contribute. With $\vec{q}\parallel [110]$ and $\vec{u} \parallel [001]$, sound waves generate equal $S_4$ and $S_5$ components of the strain tensor, as shown in Fig.~\ref{sample}(c). For the $4mm$ point group of TaAs, the $d_{24} = d_{15}$ components of the piezoelectric tensor are non-zero~\cite{buckeridge2016,dieulesaint}, thus through the relation $P_i = d_{ij}S_i$, a dielectric polarization is generated parallel to $\vec{q}$. In the absence of conduction electrons, this leads to an additional restoring force which increases the sound velocity relative to the bare sound velocity $v_0$ in the absence of piezoelectricity. However, conduction electrons can screen this dielectric polarization, reducing the effect.  As discussed in Refs.~\cite{hutson1962,rinkel2019} for example, one can obtain, for this particular mode, the relation
\begin{equation}
\omega^2 = \vec{q}^2 \left[ v_0^2 + \frac{d_{15}^2}{\rho (\epsilon_\infty + i q_j \sigma_{jk}q_k/\omega |\vec{q}|^2)}\right].
\label{sigma_to_omega}
\end{equation}
The sound velocity is then determined with $v_{44} = \omega / \mathcal{R}(|\vec{q}|)$. While in principle this provides us with a quantitative relationship between sound velocity and conductivity $\vec{\sigma}$, in practice it is difficult to obtain precise values of the parameters used in Eq.~\ref{sigma_to_omega}. Qualitatively speaking, an \emph{increase} in conductivity along the direction of sound wave propagation will lead to a \emph{decrease} in sound velocity.

The measurements were carried out using a single axis rotator to change the orientation of the magnetic field with respect to the crystalline axes and therefore the sound propagation direction $\vec{q}$. We define $\theta$ as the angle between the field and the $c$-axis direction (see Fig.~\ref{sample}a) and $\varphi$ as the angle between the field and the sound propagation direction $\vec{q}\parallel [110]$ in the $\theta=90^\circ$ plane. Since the crystal structure is tetragonal, we expect fairly significant anisotropy as we turn the field from $\theta=90^\circ$ to $0^\circ$. However, if there are anisotropies as a function of $\varphi$ in the $\theta=90^\circ$ plane (\ie~a breaking of the $C_4$ symmetry by magnetic field), they may originate from the chiral anomaly. For instance, in the absence of the chiral anomaly we would expect very little difference between $\varphi = 0^\circ$ (that is $\vec{B}\parallel\vec{q}\parallel [110]$) and $\varphi = 90^\circ$ (that is $\vec{B} \parallel [1\bar{1}0] \perp \vec{q}$). Both of these directions are crystallographically equivalent and, moreover, the projection of all of the various Fermi surfaces [a quarter of which are shown in Fig.~\ref{sample}(b)] along the magnetic field is identical. This would not be the case for sound wave propagation along $[100]$ for example. That said, the application of magnetic field within the $ab$-plane could break the $C_4$ symmetry of the system in other ways and definitive proof of a topologically non-trivial effect will ultimately depend on a connection between realistic theory and experiment. Presumably the same dilemma applies to most other experimental techniques, including transport measurements. 


 \begin{figure}[t]
 \centering
  \includegraphics[width=8cm]{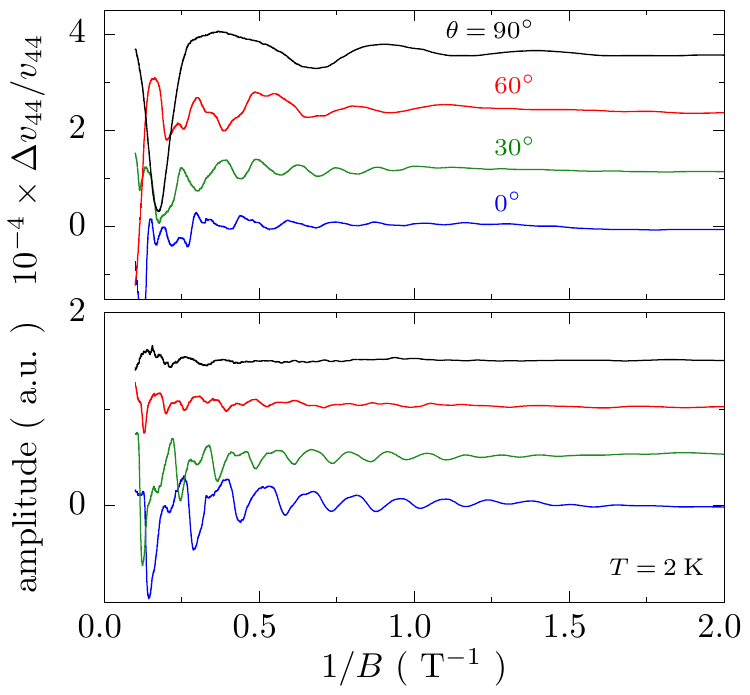}
 \caption{(Top) Field dependence of the sound velocity at $T=2$~K for various orientations of the
 magnetic field with respect to the ultrasound propagation, from $\theta=0$,
 ($\vec{B} \parallel [001]$, which is perpendicular to $\vec{q}$) to $\theta=90^\circ$
 ($\vec{B} \parallel [110]$, which is parallel with $\vec{q}$), plotted as a function of $1/B$.
 (Bottom) Same plot for the amplitude. For all of the curves, $\varphi = 0$. The curves were shifted vertically for clarity.
 }
 \label{theta}
 \end{figure}



 \begin{figure*}
 \centering
 \includegraphics{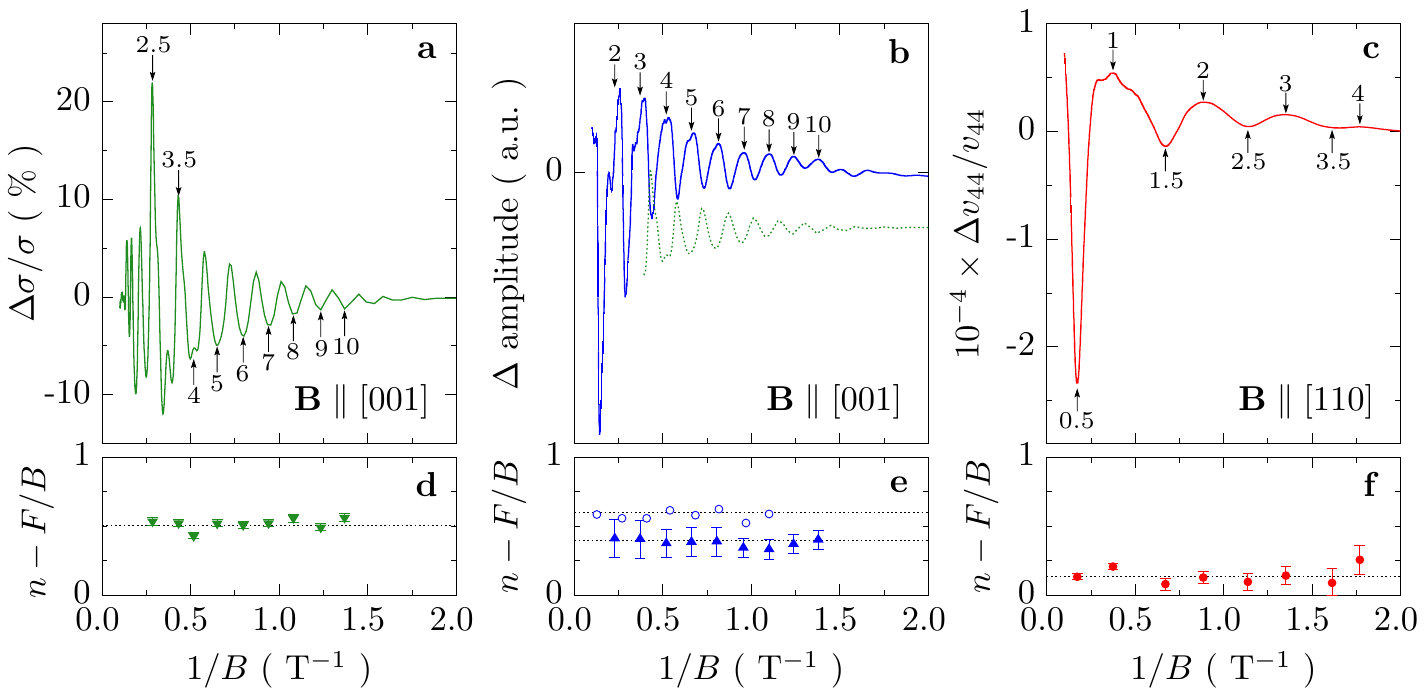}
 \caption{{\bf Upper panels:} QOs for three measured quantities at $T=2$~K and
for two distinct directions of the magnetic field, revealing two QO frequencies (2 and 7~T).
(a) SdH effect form the W1 pocket at $\theta = 0$ on a similar sample to the one used for ultrasound measurements. The irregular shapes of maxima at low $1/B$ arise from the second harmonic.
(b) Oscillations from the W1 pocket in ultrasound amplitude at $\theta=0$ (blue curve), compared with the SdH oscillations (dashed green curve).
(c) Oscillations in sound velocity, $\Delta v/v$, from the H1 pocket at $\varphi=0$ and $\theta=90^\circ$.
{\bf Lower panels:} LL indices as a function of $1/B$, from the values identified in the upper panels, plotted as $n-F/B$ where $F$ is the main QO frequency obtained from a linear fit of $n$ vs $1/B$. (d) For this pocket, the intercept is compatible with $\Phi_\mathrm{B} = \pi$. 
(e) LL indices from sound amplitude (triangles) and sound velocity (circles) give a non-trivial Berry phase as in (d) provided we identify maxima as integer indices. There is an approximate $\pi/2$ phase shift between sound velocity and amplitude oscillations, as expected. 
(f) Keeping the same convention for minima and maxima, we find an intercept of 0.135 for the 2~T oscillations in velocity. This is roughly a $\pi$ phase shift with respect to the velocity oscillations arising from the Weyl pocket and is therefore compatible with a trivial Berry phase for this pocket.
}
 \label{QOs}
 \end{figure*}


\section{Results}

The presentation of our results is organized into two sections. First we discuss the observed QOs and their implications for the topological nature of the Fermi surfaces of TaAs shown in Fig. 1b. Next we discuss the in-plane anisotropy (as a function of $\varphi$) as a possible demonstration of the chiral anomaly.


\subsection{ Quantum oscillations}

 While the quantum oscillations (QOs) observed in our measurements (see Fig.~\ref{theta}) create a complicated background signal for possible signatures of the chiral anomaly, they also provide an opportunity to study the various Fermi surfaces that are coupled to the lattice. The Fermi surface topology in TaAs (see Fig.~\ref{sample}b) has previously been investigated by Arnold~\etal~~\cite{arnold2016a} by means of angle-dependent measurements of quantum oscillations in magnetization, magnetic torque and magneto-resistance, providing a useful point of comparison for our results. The ultrasound measurements presented here similarly show clear oscillations periodic in $1/B$, as shown in Fig.~\ref{theta}, for both the velocity and the echo amplitude (which is related to the inverse of attenuation).

A fast Fourier transform (FFT) performed for $\theta = 0$ gives a dominant frequency of 6.9~T, in good agreement with
that found in Ref.~\cite{arnold2016a} for $ \vec{B} \parallel [001]$, which was attributed to a set of electron
pockets (W1, orbit $\alpha$) each containing a Weyl node. The temperature evolution of the FFT amplitude confirms the very light effective mass (of order 1\% of the free electron mass) expected in such a material. As the angle $\theta$ is moved away from 0, the frequency increases as $1/\cos\theta$, again in agreement with Ref.~\cite{arnold2016a}. For $\theta=90^\circ$, the amplitude of oscillations from the W1 pocket is negligible and a smaller frequency of 2.1~T is dominant. This is close to the value obtained by Arnold~\etal~
for $\vec{B}\parallel [110]$ which was attributed to a topologically trivial hole pocket (H1, orbit $\beta$).

In order to confirm these Fermi surface assignments, we consider the relative Berry phase between the two distinct pockets to which we are coupled. In Fig.~\ref{QOs}a and Fig.~\ref{QOs}b, we focus on the data at $\theta=0$ which most clearly show oscillations from the W1 pocket. In Fig.~\ref{QOs}c, we present data for $\theta=90^\circ$ and $\varphi=0$ highlighting the oscillations from the H1 pocket. Each QO extremum is the result of the crossing of a LL and the chemical potential. As explained in Ref.~\cite{doiron-leyraud2015}, the conductivity $\sigma_{xx}$ is minimal when an integer number $n$ of LLs is filled. The field positions of these minima, $B_n$, are then described by the following equation
\begin{align}
     \left(\frac{F}{B_n} - \delta + \frac{\Phi_\mathrm{B}}{2\pi} \right) = n-\frac 12 \>.
     \label{berry}
\end{align}
The left-hand side contains the frequency $F$, the Maslov index $\delta = 1/2 + \gamma$ (with $\gamma=0$ in two dimensions and $\gamma=\pm1/8$ in three dimensions, where the sign is given by the maximal or minimal cross section of the Fermi surface~\cite{Lukyanchuk}), and the Berry phase $\Phi_{\rm B}$. No additional phase difference is expected from the hole-like and electron-like nature of the pockets for the SdH effect~\cite{Lukyanchuk}. Neglecting temporarily $\gamma$ in Eq.~\ref{berry}, this reduces to $F/B + \Phi_\mathrm{B}/2\pi=n$. Hence a plot of $n$ as a function of $1/B$ (known as an Onsager plot) directly gives the frequency $F$ as the slope and the Berry phase $\Phi_B/2\pi$ as the intercept. Similarly, in the lower panels of Fig.~\ref{QOs}, we have plotted $n-F/B$ vs $1/B$, using the fitted value of $F$, revealing a constant value equal to $\Phi_\mathrm{B}/2\pi$.

In order to properly identify the extrema, we first compare the QOs obtained in the conductivity $\sigma_{xx}$ (Fig.~\ref{QOs}a, d) and ultrasound measurements (Fig.~\ref{QOs}b, e) for the same field orientation, that is with $\theta=0$. The conductivity was measured on a different sample from the same growth. An Onsager plot of conductivity (minima as integer indices) for this configuration yields a slope of 7~T and an intercept of 0.5, implying $\Phi_{\rm B} = \pi$ (again as long as we take $\gamma=0$). A direct comparison of the  extrema shows that minima of conductivity correspond roughly to maxima in sound velocity. Therefore, integer indices are attributed to maxima and used in the Onsager plot in Fig.~\ref{QOs}e (blue circles). More precisely, the intercept in Fig.~\ref{QOs}e for the sound velocity is $0.58 \pm 0.03$. Again, this likely implies $\Phi_{\rm B} = \pi$ and would allow for a non-zero value of $\gamma$. Oscillations in signal amplitude (shown in Fig.~\ref{QOs}b) are phase shifted by roughly $\pi/2$ with respect to the sound velocity, as can be seen from the change in intercept in Fig.~\ref{QOs}e (blue triangles). This is expected given that the sound velocity and attenuation represent the real and imaginary parts of the acoustic phonon dispersion relation, respectively.

Moving on to the sound velocity of the H1 pocket ($\beta$-orbit) measured at $\theta=90^\circ$, shown in Fig.~\ref{QOs}c, and applying the same identification of maxima with integer values of $n$, we see that the intercept in Fig.~\ref{QOs}f is roughly 0.135. We can see that the difference in $\Phi_{\rm B}/2\pi$ between the H1 and W1 pockets is 0.465, that is very close to 1/2. Hence, our quantum oscillation measurements confirm the conclusions of Arnold \emph{et al.}~\cite{arnold2016a}, showing that the 7 T QO frequency (from the W1 pocket) is topological in nature whereas the 2 T QO frequency (from the H1 pocket) originates from a trivial hole pocket. The 0.135 value of the offset in Fig.~\ref{QOs}f, may imply $\gamma=1/8$, as expected for a 3-dimensional band structure though it remains unclear why this additional offset does not appear in the conductivity measurements.

It is worth emphasizing that, for the H1 trivial hole pocket, the last LL crosses the chemical potential around 6~T, leaving the electronic structure in the quantum limit, which means that all electrons in this pocket are confined to the highly degenerate $n=0$ LL. The magnetic fields employed here are not, however, sufficient to reach the final chiral LL of the Weyl pockets.


%
\subsection{High-field anisotropy} 
\begin{figure*}
 \centering
 \includegraphics[width=6in]{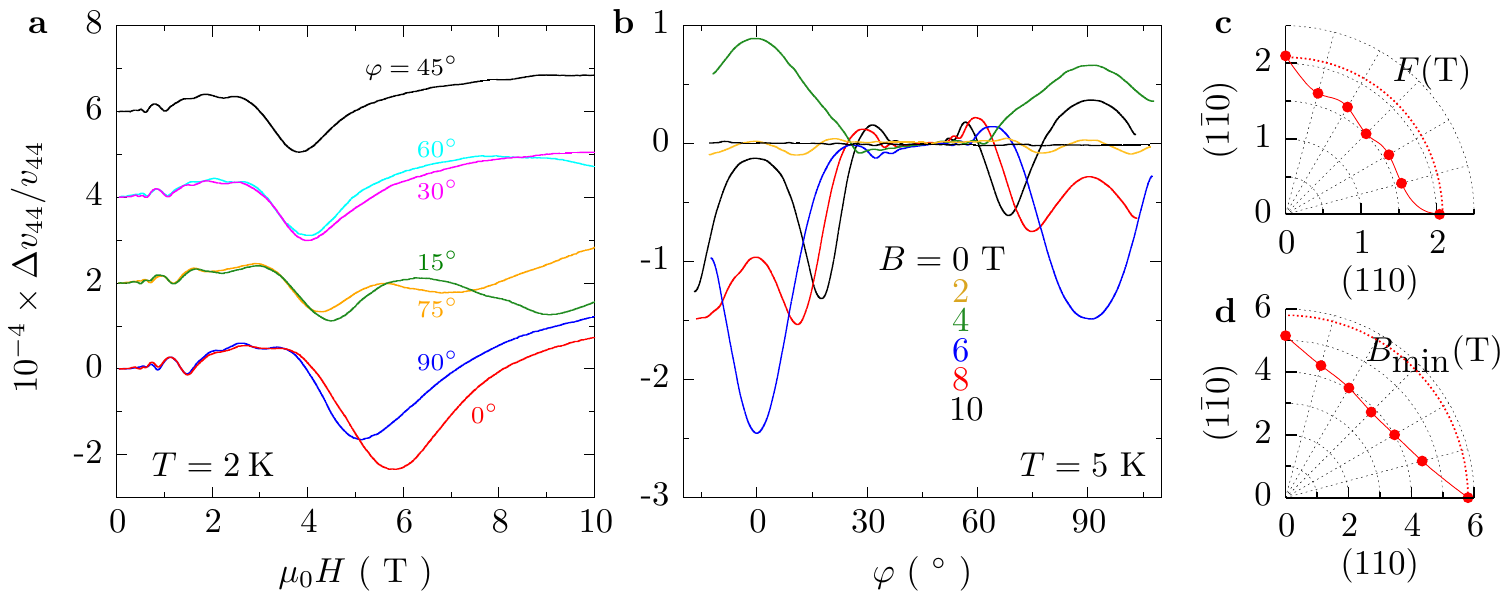} 
 \caption{(a) Field dependence of the sound velocity at $T=2$~K for various orientations of
 the magnetic field with respect to the ultrasound propagation, from $\varphi=0$
 (${\bf q} \parallel {\bf B} \parallel 110$) to $\varphi=90^\circ$
 (${\bf q} \perp {\bf B} \parallel 1\bar{1}0$). The curves were shifted vertically for clarity.
 (b) Angle dependence of the sound velocity at $T=5$~K for different values of the magnetic field. The field is rotated within the (001) plane, where lies the propagation of the sound waves. The curves are plotted with respect to their values at $\varphi=45^\circ$. (c) Angular variation of the quantum oscillation frequency $F$, showing $C_4$ symmetry. (d) Angular variation of the field value at which the sound velocity is minimal, $B_\mathrm{min}$, showing a breaking of the $C_4$ symmetry.}
 \label{angles}
 \end{figure*}

In order to get beyond the complicated angle-dependent QO background and search for signs of the chiral anomaly, we focus here on measurements performed with the field in the $ab$-plane where the oscillations are not as strongly angle dependent. Our main results are shown in Fig.~\ref{angles} in two contrasting ways. First, in Fig.~\ref{angles}(a), we plot the relative change in sound velocity $(v_{44}(B)-v_{44}(0))/v_{44}(0)$ as a function of magnetic field up to 10~T, at $T=2$~K, and for different values of the angle $\varphi$ between the sound propagation direction and the applied magnetic field. Above a threshold field that happens to be close to the last minimum at $\sim5$~T, a significant anisotropy appears. The difference between $\varphi=0$ and $90^\circ$ curves is particularly informative as these are crystallographically equivalent field orientations and one would naively expect identical results. This appears to be the case for the low-field QOs as seen in the polar plot of Fig.~\ref{angles}(c), where the QO frequency respects the tetragonal symmetry of the lattice. However, the difference between $\varphi=0$ and $90^\circ$ curves grows rapidly from 5 to 6~T before slightly decreasing to remain at a constant value above roughly 8~T. 

In Fig.~\ref{angles}(b), the sound velocity is measured as a function of the angle $\varphi$ for various values of $B$ from 0 to 10~T at $T=5$~K, again revealing anisotropy with respect to field direction. A polar plot in Fig.~\ref{angles}(d) of the field value at which the sound velocity is minimal, $B_\mathrm{min}$, shows a clear breaking of the $C_4$ symmetry of the $ab$-plane, again contrasting with the QO frequency (Fig.~\ref{angles}c). Note that a slight misalignment of the sample away from $\theta=90^\circ$ cannot explain this behavior and would instead result in an observable change in the QO frequency. 

Another well-known mechanism for field-induced anisotropy in sound velocity is the Alpher-Rubin effect, whereby the Lorentz force on electronic currents generated by the oscillating ionic charges leads to an \emph{increase} in the velocity of transverse sound waves when $\vec{B}$ is parallel with $\vec{q}$~\cite{Luthi,Rodriguez1963}. However this effect is found (based on our sample's conductivity of $\sigma_{xx} \simeq 1\times 10^4$ $\Omega^{-1}$m$^{-1}$ at 10 T) to contribute a negligible change in velocity ($\Delta v / v \sim10^{-10}$) and, moreover, is at odds with the observed \emph{decrease} in sound velocity when $\vec{B}\parallel\vec{q}$. 

Hence, we attribute this roughly 50 ppm breaking of tetragonal symmetry in the sound velocity at high magnetic fields to an anisotropy in the screening of the strain-induced dielectric polarization resulting from an anisotropy in conductivity. In this regime we are in the quantum limit of the topologically trivial hole pockets and the Weyl nodes may provide the most important contribution to changes in sound velocity. We propose that since sound waves generate an oscillating polarization parallel to $\vec{q}$ along with screening currents which are more effective when $\vec{B}\parallel\vec{q}$, the observed anisotropy is likely a demonstration of LNMR caused by the chiral anomaly.

As predicted by Rinkel \emph{et al.}~\cite{rinkel2019}, the difference in velocity (between $\varphi=0^\circ$ and $\varphi = 90^\circ$) reaches a constant value at high field, and the velocity is found to be reduced for $\varphi=0^\circ$ (see Fig.~\ref{angles}a). This is explained by the fact that, in the absence of conduction electrons, piezoelectric coupling leads to an increased velocity. Higher conductivity leads to better screening of the dielectric polarization and therefore a drop in sound velocity toward the value it would take in the absence of piezoelectricity. Hence, the drop in sound velocity here could possibly be attributed to LNMR. Ref.~\cite{rinkel2019} predicts such a drop in velocity to occur for all angles of the magnetic field, except for a narrow window around $\varphi = 90^\circ$. However, it is also noted that this window of increased sound velocity would become broader and more easily observable once sample disorder is considered. We cannot carefully study the angular dependence of this difference in velocity given the complexity of the underlying QOs and their dependence on $\varphi$. In principle, a better approach would be to maintain a fixed field angle and vary the angle of sound-wave propagation, but this would be prohibitively difficult. 

It should also be noted that there is a considerable discrepancy (well beyond the measurement uncertainty) between the measured anisotropy (50 ppm) and the theory of Ref.~\cite{rinkel2019} (nearly 40\%). The calculation of Ref.~\cite{rinkel2019} is based on realistic parameters for TaAs, determined from experiment or \emph{ab initio} calculations~\cite{buckeridge2016}, but does consider only the contribution from the Weyl fermions, which are modelled through a spherically symmetric Hamiltonian. Furthermore, since the predicted angular variation of sound velocity comes only from the lowest energy chiral LL, the effect may be heavily diluted by other LLs when not in the quantum limit of the Weyl nodes, as is the case in our experiments. 

Evidently more detailed theoretical calculations would be valuable for understanding whether the measured effect can be entirely attributed to the topological nature of the Weyl nodes. Since the application of a magnetic field in the $ab$-plane necessarily breaks the $C_4$ symmetry of the lattice in any tetragonal system, and therefore could generate an anisotropic sound velocity through a different mechanism, it is crucial to quantify the effect of the chiral anomaly. Similarly, a more robust understanding of this phenomenon could also be achieved with a campaign of similar experiments on a variety of topological and trivial semimetallic systems. A study of NbAs could be particularly beneficial. While it has similar structure and band structure to TaAs, \emph{ab initio} simulations conclude that the Weyl nodes are well below the Fermi energy and nodes of opposing chirality are contained within the same Fermi surface pockets meaning that the chiral anomaly should not be present~\cite{Lee2015}.

\section{Conclusion}

To summarize, we have carried out sound velocity measurements on the Weyl semimetal TaAs, as a function of magnetic field and field angle with respect to the sound propagation direction. The observed quantum oscillations are found to be consistent with the dHvA and SdH measurements of Arnold~\emph{et al.}~\cite{arnold2016a} and the phase of these oscillations have allowed us to identify a non-zero Berry phase for one of the topological Weyl pockets. With the field angle varied in the $ab$-plane, a significant anisotropy that breaks the $C_4$ symmetry of the structure is observed at relatively high field (above $\sim 5$ T). This anisotropy is qualitatively consistent with theoretical predictions~\cite{rinkel2019} and might, therefore, be attributed to the chiral anomaly, essentially providing a measurement of the negative longitudinal magnetoresistance without electrical contacts and the extrinsic current-jetting effects that result. 

\begin{acknowledgements}
We are grateful to M. Castonguay for extensive technical support, and we acknowledge valuable discussions with A. Burkov, N. Doiron-Leyraud, D. LeBoeuf, G. Quirion, C. Proust, B. J. Ramshaw and E. Hassinger. In particular, we are grateful to I.~Garate and P.~Rinkel who kept us apprised of their theoretical work on this same subject. Work at the Universit\'{e} de Sherbrooke was supported by the National Sciences and Engineering Research Council (NSERC), the Fonds de recherche du Qu\'ebec - Nature et technologies (FRQNT) and the Canada First Research Excellence Fund (CFREF). Experiments were performed using equipment funded by the Canadian Foundation for Innovation (CFI). GA and NN acknowledge support from the National Science Foundation under Grant No. 1607753.
\end{acknowledgements}



\end{document}